# A combinatorial problem about binary necklaces and attractors of Boolean automata networks

Mathilde Noual[1]

Aix-Marseille Université, CNRS, LIF UMR 7279, 13288, Marseille, France

**Abstract.** It is known that there are no more Lyndon words of length $n$ than there are periodic necklaces of same length. This paper considers a similar problem where, additionally, the necklaces must be without some forbidden factors. This problem relates to a different context, concerned with the behaviours of particular discrete dynamical systems, namely, Boolean automata networks. A formal argument supporting the following idea is provided: *addition of cycle intersections in network structures causes exponential reduction of the networks' number of attractors.*

## *1 Introduction and informal motivation*

Generally, I aim at understanding the "clockworks" of interaction networks (*a.k.a. sets of related things*) through a study of formal prototypes called *Boolean automata networks* (BANs). These mathematical models are a generalisation of the neural networks introduced by McCulloch and Pitts [15] in 1943. They are still widely studied as models of biological networks. Considerable effort has thereby already been invested into understanding and describing their dynamics[1] [2, 5, 9, 16, 21, 27, 31].

More precisely and informally, here, I especially aim at understanding what structural properties can be considered fundamentally responsible for *diversity* and *variety* in the asymptotic behaviour of a network, and conversely, which ones can be considered responsible for a lesser degree of "asymptotic freedom" (*cf.* Section 7). This aim leads to addressing some crucial lingering problems about BANs through a new, elementary stance. In particular, this raises a combinatorial problem about the asymptotic dynamics of particular instances of BANs (*cf.* Section 8) which translates directly into a combinatorial word theoretic problem (*cf.* Section 10).

Before describing these two equivalent combinatorial problems and how they relate (in Sections 6 to 10), in Sections 2 to 5, we give basic definitions about BANs, their structures, and their (asymptotic) behaviours. Section 11 sketches the (lengthy) proof of the main result in the word theoretic setting (the full proof is detailed in Appendix A). Finally, Sections 12 and 13 derive and discuss implications of this result.

---

[1] Commonly, BANs are studied through their dynamics, often with the aim of relating their dynamical properties to their other (structural) properties. Notably, my present aim doesn't exactly coincide with this. It is based on a general approach that is essentially constructive. To start, I *isolate* features of networks (*e.g.* cycles, intersections, non-monotony) in order to study their role in conditions that favour their decisiveness. Then, the complexity of problems addressed can gradually be increased by adding and combining features that have already been studied separately.

## 2 Boolean Automata Networks

Let $\mathbb{B} = \{0, 1\}$. A **Boolean automata network** (BAN) of size $n \in \mathbb{N}$ is a set of $n$ Boolean functions $\mathcal{N} = \{f_i : \mathbb{B}^n \to \mathbb{B}, \; i < n\}$. Index $i < n$ represents **automaton** $i$. Here, the word *automaton* is to be taken as referring to a computing unit regarded as a *black box* (our interest here is in how *networks* of automata work, rather than in how automata work) [4]. The computation that is made by automaton $i$ in **configuration** $x \in \mathbb{B}^n$ of $\mathcal{N}$ is: $x \mapsto f_i(x)$. In principle, $f_i$ can be *any* Boolean function. In practice, for the sake of convenience and in consistence with our general approach to these networks[2], we restrict the $f_i(x)$'s to *fully locally monotone functions*: in the CNF or DNF of any $f_i(x)$, no literal $x_j$ ($j < n$) can appear both negated and un-negated (typically, this excludes the XOR function).

## 3 Structure of a Boolean automata network

Let $\mathbf{V} = \{i < n\}$ denote the set of automata of $\mathcal{N}$. Interactions between automata of $\mathcal{N}$ are represented in its **interaction digraph** – also called **structure** – $\mathbf{G} = (\mathbf{V}, \mathbf{A})$, where $\mathbf{A} \subset \mathbf{V} \times \mathbf{V}$ is defined by: $(j, i) \in \mathbf{A} \Leftrightarrow \exists x = (x_0 \ldots x_n) \in \mathbb{B}^n, \; f_i(x_0 \ldots x_{j-1} 1 x_{j+1} \ldots x_{n-1}) \neq f_i(x_0 \ldots x_{j-1} 0 x_{j+1} \ldots x_{n-1})$. In $\mathbf{G}$, arc $(j, i)$ is said to be **negative** (resp. **positive**) if:

$$\forall x \in \mathbb{B}^n, \quad f_i(x_0 \ldots x_{j-1} 1 x_{j+1} \ldots x_{n-1}) \quad \leq \quad (\text{resp.} \geq) \quad f_i(x_0 \ldots x_{j-1} 0 x_{j+1} \ldots x_{n-1}).$$

Because of the assumption on the local monotony of the $f_i$'s, all arcs can be signed. Naturally, we let $s_{j,i} \in \{+, -\}$ denote the **sign** of arc $(j, i) \in \mathbf{A}$. Then, $\forall x \in \mathbb{B}^n, \forall (j, i) \in \mathbf{A}$, we can introduce the following notation to denote the input that $i$ receives from $j$ in configuration $x$: $s_{j,i}(x_j) = x_j$ if $s_{j,i} = +$ and $s_{j,i}(x_j) = \neg x_j$ if $s_{j,i} = -$.

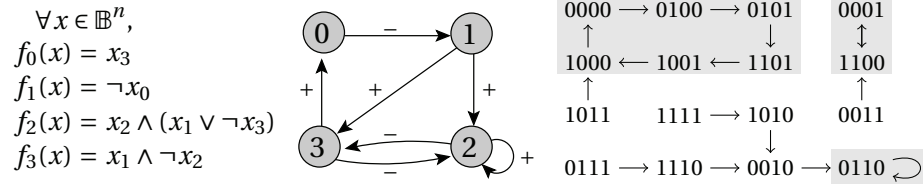

$\forall x \in \mathbb{B}^n,$
$f_0(x) = x_3$
$f_1(x) = \neg x_0$
$f_2(x) = x_2 \wedge (x_1 \vee \neg x_3)$
$f_3(x) = x_1 \wedge \neg x_2$

**Fig. 1.** A Boolean automata network $\mathcal{N}$ of size $n = 4$. Left: The defining local functions of $\mathcal{N} = \{f_i, i < n\}$. Middle: The *signed* version of the structure $\mathbf{G}$ of $\mathcal{N}$. Right: The transition graph $\mathcal{T}$ of $\mathcal{N}$, *i.e.* the graph of function $F : x \in \mathbb{B}^4 \mapsto (f_0(x), \ldots, f_3(x)) \in \mathbb{B}^4$ revealing one 1-attractor, one 2-attractor and one 6-attractor so that the order of $\mathcal{N}$ is $\omega = \text{lcm}\{1, 3, 6\} = 6$.

All walks and cycles mentioned in this paper are considered to be directed. The sign of a walk or cycle is the product of the signs of its arcs: a negative (resp. positive) walk or cycle in $\mathbf{G}$ is comprised of an *odd* (resp. *even*) number of negative arcs.

---

[2] We take (local) monotony as a reference and aim at understanding monotone BANs first so that we can then aim at understanding the role of non-monotony *per se* by studying how a little, localised addition of it impacts on the network's behaviour.

## 4 Behaviour of a Boolean automata network

Assuming a parallel update of each automaton state in each network configuration, $\mathcal{N}$ undergoes transitions of the form $x \to F(x) = (f_0(x), f_1(x),\ldots, f_{n-1}(x))$ (as each automaton $i$ undergoes change $x_i \to f_i(x)$). Given a configuration $x \in \mathbb{B}^n$, if we settle that $x = x(0)$, then $\forall t \in \mathbb{N}$, $x(t)$ denotes configuration $F^t(x)$. The graph $\mathcal{T}$ of function $F$ is called the **transition graph** of $\mathcal{N}$. It represents the *behaviour* of $\mathcal{N}$ under the parallel updating.

## 5 Asymptotic behaviour of a Boolean automata network

In the present deterministic case, terminal strongly connected components of $\mathcal{T}$ are directed cycles. To avoid confusion with the structural cycles of **G**, a cycle of length $p$ in $\mathcal{T}$ is rather called an **attractor of primitive period p** or a **p-attractor** (abusing language since an attractor need not attract anything in this setting).

We introduce the **order** $\omega$ of $\mathcal{N}$ as the least common multiple of all of its attractor periods. Equivalently, with $\mathbf{X} \subset \mathbb{B}^n$ denoting the set of **recurrent configurations** of $\mathcal{N}$ (those belonging to its attractors), $\omega$ is defined by: $\omega = \min\{p \in \mathbb{N},\ \forall x \in \mathbf{X},\ F^p(x) = x\}$.

We let $\mathbf{X}(p) = \{x \in \mathbb{B}^n,\ F^p(x) = x\}$ denote the set of configurations of period $p$. In particular, $\forall p \in \mathbb{N}, \mathbf{X}(p) \subset \mathbf{X}(\omega) = \mathbf{X}$, and $\mathbf{X}(p) \neq \emptyset \implies p | \omega$. Let us introduce here the following notation: $\forall x \in \mathbf{X},\ \forall t \in \mathbb{Z},\ x(t) = F^{t \bmod \omega}(x) = x(t \bmod \omega)$.

The **primitive period** of any $x \in \mathbf{X}$, is $\min\{p,\ F^p(x) = x\}$. We let $\widetilde{\mathbf{X}}(p)$ be the set of configurations with *primitive* period $p$: $\widetilde{\mathbf{X}}(p) = \mathbf{X}(p) \setminus \bigcup_{q|p,\ q<p} \mathbf{X}(q)$.

Let us abuse language and notations to confuse attractors with orbits $\{F^k(x),\ k < p\}$ of configurations $x$ inducing them. We let $\widetilde{\mathbf{A}}(p) = \bigcup_{x \in \widetilde{\mathbf{X}}(p)} \{\{F^k(x),\ k < p\}\}$ denote the set of $p$-attractors of $\mathcal{N}$, and we let $\mathbf{A}(p) = \bigcup_{q|p} \widetilde{\mathbf{A}}(q)$ denote its set of attractors with period $p$. In particular, $\mathbf{A}(\omega)$ is the set of all attractors of $\mathcal{N}$.

## 6 Preliminary combinatorial notations and relations

Let us specify notations for cardinals of the sets introduced above: $\forall p,\ \mathtt{X}(p) = |\mathbf{X}(p)|$, $\widetilde{\mathtt{X}}(p) = |\widetilde{\mathbf{X}}(p)|$, $\mathtt{A}(p) = |\mathbf{A}(p)|$, and $\widetilde{\mathtt{A}}(p) = |\widetilde{\mathbf{A}}(p)|$.

Provided a characterisation of attractor periods yielding $\omega$, and a characterisation of $\mathbf{X}$ yielding $\mathtt{X}(\omega)$, one can immediately derive $\widetilde{\mathtt{X}}(p)$, $\widetilde{\mathtt{A}}(p)$ and $\mathtt{A}(p)$, by exploiting the following relationships, where $\star$ is the Dirichlet convolution operator, $\mathbb{1}: n \in \mathbb{N} \mapsto 1$, $\mathrm{inv}: n \in \mathbb{N} \mapsto \frac{1}{n}$, $\mu$ is the Möbius function, and $\phi$ is the Euler totient:

$$\mathtt{X} = \widetilde{\mathtt{X}} \star \mathbb{1} \qquad \widetilde{\mathtt{X}} = \mathtt{X} \star \mu \qquad \widetilde{\mathtt{A}} = \mathrm{inv} \times (\mathtt{X} \star \mu) \qquad \mathtt{A} = \widetilde{\mathtt{A}} \star \mathbb{1} = \mathrm{inv} \times (\mathtt{X} \star \phi). \quad (1)$$

The 3rd relation above corresponds to the *Witt formula* counting the number of Lyndon words [3, 8, 11, 13, 14]. The last equality comes from Burnside's orbit-counting Lemma.

Let us note that the total number of attractors of a BAN is never greater than what it would be if all attractors had the largest possible period ($\omega$). Thus:

$$\mathtt{A}(\omega) \geq \mathtt{X}(\omega)/\omega. \quad (2)$$

## 7 Cycles, tangent cycles, and a more formal motivation

Our general, informal motivation described in the introduction leads to taking (formal) interest in the order $\omega$ (intuitively accounting for a form of "asymptotic diversity"), in the distributions of a network's configuration and attractor periods, and in the total number of attractors $\mathtt{A}(\omega)$ (intuitively accounting for a form of "asymptotic variety"). More precisely, we are interested in how all of these relate to the cycles in **G**, to their signs, and to their interactions.

It is commonly accepted and has been supported by formal arguments in several frameworks more or less related to BANs that cycles in the structure **G** of an interaction network $\mathcal{N}$ decisively impact on its (asymptotic) behaviour [20, 21, 23, 25, 30]. Having had so much attention, cycles are now rather well understood. The specific way that cycle intersections *per se* impact on the overall network behavioural possibilities, however, is not at all. Our need to increment understanding of cycles with some primary insight on this, drives us to taking interest in "tangent cycles".

We call **BAC** (Boolean Automata Cycle) any BAN that is structured as a simple cycle (*cf.* Table 1). We call **BAD** (Boolean Automata Double-cycle) a BAN structured as two tangent cycles (*cf.* Table 1). There are 2 types of BACs and 3 types of BADs (*cf.* Table 1). In [6, 17], the (asymptotic) behaviours (as defined in Section 5) of all these 5 types of BANs has been characterised, and explicit formulae have been derived for all the quantities introduced in Section 6 relative to them. These results are based (non-exclusively) on results stating that in all five cases *attractor periods divide positive cycle lengths without dividing negative cycle lengths*, on results summed up in Table 1, and on some results that can be derived from them using (1).

## 8 The combinatorial problem relative to Boolean automata networks

Through its implications (*cf.* Sections 12 and 13), our main result falls in line with motivations presented above. In this theorem, the lower bound of (3) follows from (2).

**Theorem 1.** *The total number of attractors of any* BAC *and almost*[3] *any* BAD *of order $\omega$ satisfies:*

$$\mathtt{X}(\omega)/\omega \;\leq\; \mathtt{A}(\omega) \leq 2 \cdot \widetilde{\mathtt{A}}(\omega) \;=\; 2 \cdot \widetilde{\mathtt{X}}(\omega)/\omega \;\leq\; 2 \cdot \mathtt{X}(\omega)/\omega. \tag{3}$$

*The least upper bound of* (3) *equivalently means that the expected value of an attractor period is big:* $\sum_{p \mid \omega,\, \widetilde{\mathbf{A}}(p) \neq \emptyset} p \cdot \widetilde{\mathtt{A}}(p) \geq \omega/2$. *Thus, almost all periodic configurations of* BACs *and* BADs *have the greatest possible primitive period ($\omega$) and* $\mathtt{A}(\omega) = \Theta(\mathtt{X}(\omega)/\omega)$.

In the case of BACs, the set of $p$-attractors $\widetilde{\mathbf{A}}(p)$ is isomorphic to the set of (unlabelled) Lyndon words of length $p$ [3, 7, 8, 10–14, 26, 29]. The existence of an injective map $\bigcup_{p \mid \omega,\, p < \omega} \widetilde{\mathbf{A}}(p) \to \widetilde{\mathbf{A}}(\omega)$ [24], implies that BACs satisfy: $\mathtt{A}(\omega) = \sum_{p \mid \omega,\, p < \omega} \widetilde{\mathtt{A}}(p) + \widetilde{\mathtt{A}}(\omega) \leq 2\widetilde{\mathtt{A}}(\omega) \leq 2\widetilde{\mathtt{X}}(\omega)/\omega$ and thus (3). Moreover, [17] proves that positive BADs behave as positive BACs of same order: the asymptotic (strongly connected) part of their transition graphs are isomorphic. Equation (3) therefore holds for all BACs and positive BADs.

---

[3] As the detailed proof in Appendix A reveals, with the notations introduced further on in this paper, Theorem 1 holds for all BADs except those satisfying $(K = 10 \wedge \Delta = 1) \vee (K = 6 \wedge \Delta = 2)$. That is, Theorem 1 holds for all BADs except the 3 types of negative BADs such that either $(\ell, r)$ or $(r, \ell)$ belongs to $\{(1, 9), (3, 7), (2, 10)\}$.

**Table 1.** Summary of results concerning the behaviour of BACs and BADs. $\forall k, m \in \mathbb{N}$, $\neg(k|m)$ equals 0 if $k|m$ and 1 otherwise. For BADs, $\Delta = \gcd(\omega, \ell)$, $K = \omega/\Delta$, $\Delta_p = \gcd(p, \ell) = \gcd(p, d) = \gcd(\Delta, p)$ and $K_p = p/\Delta_p$. $(L(n))_{n \in \mathbb{N}}$ is the **Lucas Sequence** [19, 22] (sequence A204 of the OEIS [28]) and $(P(n))_{n \in \mathbb{N}}$ is the **Perrin Sequence** [1] (sequence A1608 of the OEIS).

| Type of network $\mathcal{N}$ | **BAC of size n** (network $\mathcal{N}$ whose interaction graph **G** is a simple cycle of length $n$) | | **BAD** of *left-length* $\ell$, of *right-length* $r$, and of size $n = \ell + r - 1$ ($\mathcal{N}$ s.t. **G** is isomorphic to two tangent cycles of respective lengths $\ell$ and $r$) | | |
|---|---|---|---|---|---|
| | **Positive BAC** (BAC s.t. **G** is a cycle with an even number of negative arcs) | **Negative BAC** (BAC s.t. **G** is a cycle with an odd number of negative arcs) | **PositiveBAD** (BAD s.t. **G** consists of two tangent positive cycles) | **Mixed BAD** (BAD s.t. **G** consists of two tangent cycles of different signs) | **Negative BAD** (BAD s.t. **G** consists of two tangent negative cycles) |
| Order $\omega =$ | $n$ | $2n$ | $\gcd(\ell, r)$ | $r$, the length of the postive cycle | $\begin{cases} \ell + r & \text{if } K \neq 4 \\ \frac{\ell+r}{2} & \text{if } K = 4 \end{cases}$ |
| $\forall p \| \omega$, $\mathbf{X}(p) =$ | $2^p$ | $\neg(p\|\frac{\omega}{2}) \cdot \sqrt{2}^p$ | $2^p$ | $\neg(p\|\Delta) \cdot L(K_p)^{\Delta_p}$ | $P(K_p)^{\Delta_p}$ |
| $\mathbf{F}_d =$ | $\emptyset$ | $\{b w b \mid w \in \mathbb{B}^{d-1}, b \in \mathbb{B}\}$ | $\emptyset$ | $\{0w0 \mid w \in \mathbb{B}^{d-1}\}$ | $\{0w0, 1w1v1 \mid w, v \in \mathbb{B}^{d-1}\}$ |

The two remaining cases turned out to be considerably more tricky to deal with: in [17], we could only conjecture that (3) also holds for mixed and negative BADs also. In the sequel of this paper, we prove this conjecture.

## 9 From automata orbits to binary words without some specific forbidden factors

Although we are now going to focus on BADs, in order to clarify how things work, let us first note that in a BAC of size $n$ and sign $s$, identifying **V** with $\mathbb{Z}/n\mathbb{Z}$, we have: $\forall i \in \mathbf{V}$, $f_i(x) = s_{i-1,i}(x_{i-1})$. It follows that any configuration $x = x(t) \in \mathbb{B}^n$ satisfies $x(t+n) = x$ if $s = +$ and $x(t+n) = (\neg x_0 \ldots \neg x_{n-1})$ if $s = -$.

In a BAD $\mathcal{D}$ of size $n = \ell + r - 1$, all automata with in-degree 1 also satisfy $f_i(x) = s_{i-1,i}(x_{i-1})$ and the intersection automaton $i = 0$ satisfies: $f_0(x) = s_{\ell-1,0}(x_{\ell-1}) \diamond s_{n-1,0}(x_{n-1})$, where $\diamond \in \{\vee, \wedge\}$ (this is because of the assumed local monotony of $f_0$). We are going to concentrate on the case where $\diamond = \vee$. The case where $\diamond = \wedge$ is similar. Let $s^L$ (resp. $s^R$) be the sign of the left (resp. right) cycle of $\mathcal{D}$. Let $d = \ell \bmod p$ and $d' = r \bmod p$. Importantly, if $p$ is a network period ($\mathbf{X}(p) \neq \emptyset$), and $s^L = -$, then (*cf.* end of Section 7) $\neg(p|\ell)$ and consequently $d > 0$.

Any recurrent configuration $x = x(t) \in \mathbf{X}(p)$ of the BAD satisfies:

$$x_0(t) = \begin{cases} \neg x_0(t-d) \vee x_0(t) & \text{when } s^L = - \text{ and } s^R = + \\ \neg x_0(t-d) \vee \neg x_0(t+d) & \text{when } s^L = s^R = - \end{cases} \quad (4)$$

(because, according to the 3rd line of Table 1, $d' \bmod p = 0$ in the first case, $d' \equiv p - d \bmod p$ in the second, and in both $x_0(t) = f_0(x(t-1)) = s_{\ell-1,0}(x_{\ell-1}(t-1)) \vee s_{n-1,0}(x_{n-1}(t-1)) = s_{\ell-1,0} \circ s_{\ell-2,\ell-1}(x_{\ell-2}(t-2)) \vee s_{n-1,0} \circ s_{n-2,n-1}(x_{n-2}(t-2)) = \ldots = s^L(x_0(t-\ell)) \vee s^R(x_0(t-r)) = s^L(x_0(t-\ell \bmod p)) \vee s^R(x_0(t-r \bmod p)) = s^L(x_0(t-d)) \vee s^R(x_0(t-d')))$.

Now, consider the orbit of automaton 0 of $\mathscr{D}$ that is initiated in configuration $x = x(t)$. Let $w = x_0(t)x_0(t+1)\ldots x_0(t+p-1) \in \mathbb{B}^p$ be the word naturally defined by it. Following (4), letters of $w$ satisfy relations. To explicit this, $\forall d \in \mathbb{N}^*$ (representing $\ell \bmod p$), we define a **set of forbidden factors relative to** $d$:

$$\mathbf{F}_d = \begin{cases} \{0w0 \mid w \in \mathbb{B}^{d-1}\} & \text{if } s^L \neq s^R \\ \{0w0, 1w1v1 \mid w, v \in \mathbb{B}^{d-1}\} & \text{if } s^L = s^R = -. \end{cases}$$

We then define the following set that will play an essential role in the sequel:

$$\mathbf{W}_d^n = \{w \in \mathbb{B}^n, w \text{ contains no factor in } \mathbf{F}_d\} \subset \mathbb{B}^n. \tag{5}$$

Importantly, let us note that $\forall d > 0$, $\mathbf{W}_d^d = \emptyset$. Also, $mathbf{W}_d^n = mathbf{W}_{n-d}^n$. Thus, (4) implies part of the following Lemma which is proven in [17], and which can also be stated for BACs using the corresponding sets $\mathbf{F}_d$ defined in Table 1.

**Lemma 1.** *Let $p \in \mathbb{N}$, let $x \in \mathbf{X}(\omega)$ be a recurrent configuration of a* BAD *of order $\omega$ with negative left cycle of length $\ell \equiv d \bmod p$, and let $w \in \mathbb{B}^\omega$ be the orbit of intersection automaton* 0 *that is initiated in $x$. Then:*

$$x \in \mathbf{X}(p) \Leftrightarrow w \in \mathbf{W}_d^p$$

*(where $\mathbf{X}(p) \neq \emptyset \implies (p|\omega) \wedge \neg(p|\ell) \Leftrightarrow (p|\omega) \wedge \neg(p|\Delta) \wedge (d > 0)$).*

## 10 The combinatorial problem relative to binary necklaces

Henceforth, we concentrate on binary words $w \in \mathbb{B}^n$ of arbitrary length $n \in \mathbb{N}$, with letters indexed from 0 to $n-1$. We abuse notations so as to let $w_k$, $\forall k \in \mathbb{Z}$, refer to letter $w_k = w_{k \bmod n}$ of word $w$. A **necklace** of length $n \in \mathbb{N}$ represents an equivalence class of words under iterates of the rotation $\rho : w \mapsto w_{n-1}w_0\ldots w_{n-2}$. The necklace (conjugacy class) representing (containing) word $w \in \mathbb{B}^n$ and all of its rotations (conjugates) $\rho^k(w), k \in \mathbb{Z}/n\mathbb{Z}$ is denoted by $\langle w \rangle$ and we write $w \equiv w'$ when $\langle w \rangle = \langle w' \rangle$.

Additionally to set $\mathbf{W}_d^n$ defined in (5) above, $\forall p | n$ and $\forall d < n$, we define the following sets:

$$\mathbf{W}_d^n(p) = \{w \in \mathbf{W}_d^n \text{ has period } p\} = \{u^{\frac{n}{p}} \in \mathbf{W}_d^n, u \in \mathbb{B}^p\} = \{u^{\frac{n}{p}}, u \in \mathbf{W}_d^p\} \tag{6}$$

$$\widetilde{\mathbf{W}}_d^n(p) = \{w \in \mathbf{W}_d^n \text{ has primitive period } p\} = \mathbf{W}_d^n(p) \setminus \bigcup_{q|p, q<p} \mathbf{W}_d^n(q)$$

In particular, $\widetilde{\mathbf{W}}_d^n = \widetilde{\mathbf{W}}_d^n(n) = \{w \in \mathbf{W}_d^n \text{ is primitive (aperiodic)}\}$

$\mathbf{C}_d^n = \{\langle w \rangle, w \in \mathbf{W}_d^n\}$
$\mathbf{C}_d^n(p) = \{\langle w \rangle, w \in \mathbf{W}_d^n(p)\}$
$\widetilde{\mathbf{C}}_d^n(p) = \{\langle w \rangle, w \in \widetilde{\mathbf{W}}_d^n(p)\}$ and in particular, $\widetilde{\mathbf{C}}_d^n = \widetilde{\mathbf{C}}_d^n(n) = \{\langle w \rangle \in \mathbf{C}_d^n \text{ is primitive}\}$.

We let $\mathbf{W}_d^n$, $\mathbf{W}_d^n(p)$, $\widetilde{\mathbf{W}}_d^n(p)$, $\widetilde{\mathbf{W}}_d^n$, $\mathbf{C}_d^n$, $\mathbf{C}_d^n(p)$, $\widetilde{\mathbf{C}}_d^n(p)$, and $\widetilde{\mathbf{C}}_d^n$ straightforwardly denote the cardinals of all these sets.

The last equality of (6) holds because $w = u^{\frac{n}{p}} \in \mathbf{W}_d^n(p)$ implies that $u \in \mathbf{W}_d^p$. Consequences of this are the following, where we write $S \simeq S'$ for any sets $S$ and $S'$ with a bijective pairing between them:

$$\mathbf{W}_d^n(p) \simeq \mathbf{W}_d^p, \quad \widetilde{\mathbf{W}}_d^n(p) \simeq \widetilde{\mathbf{W}}_d^p, \quad \mathbf{C}_d^n(p) \simeq \mathbf{C}_d^p, \text{ and } \widetilde{\mathbf{C}}_d^n(p) \simeq \widetilde{\mathbf{C}}_d^p. \quad (7)$$

As a result of Lemma 1, we also have the following relations between sets of configurations and attractors (relative to a negative cycle of length $\ell \equiv d \bmod p$), and sets of words and necklaces:

$$\mathbf{X}(p) \simeq \mathbf{W}_d^p, \quad \widetilde{\mathbf{X}}(p) \simeq \widetilde{\mathbf{W}}_d^p, \quad \mathbf{A}(p) \simeq \mathbf{C}_d^p, \quad \widetilde{\mathbf{A}}(p) \simeq \widetilde{\mathbf{C}}_d^p. \quad (8)$$

As a result of (7) and (8), proving the least upper bound of Theorem 1 is equivalent to proving the following key proposition:

**Proposition 1.** *Let $n, d \in \mathbb{N}, d < n$. Let $\mathbf{F}_d$ be any of the sets of forbidden word factors defined in the last line of Table 1. The cardinals of the sets of binary necklaces defined above relatively to $\mathbf{F}_d$ almost always satisfy*[4]*:*

$$\mathbf{C}_d^n \leq 2 \cdot \widetilde{\mathbf{C}}_d^n. \quad (9)$$

In other terms, to prove that BADs, just like BACs, have no more small attractors ($p$-attractors s.t. $p < \omega$) than big ones ($\omega$-attractors), we want to show that there are no more periodic necklaces without the forbidden factors of $\mathbf{F}_d$ than there are aperiodic ones. Explicit formulae for cardinals of all sets introduced above are known [17]. However, comparing these formulae turns out to be very tricky. Thus, here, we propose to build and injective map $\gamma : \bigcup_{p \mid n, p < n} \widetilde{\mathbf{C}}_d^p \to \widetilde{\mathbf{C}}_d^n$.

The existence of this map will prove Proposition 1 and thereby Theorem 1.

## 11 Proof of the main result

We prove (9) in the case where $\mathbf{F}_d = \{0w0, 1w1v1 \mid w, v \in \mathbb{B}^{d-1}\}$, *i.e.* we prove (3) for negative BADs. The case where $\mathbf{F}_d = \{0w0 \mid w \in \mathbb{B}^{d-1}\}$ corresponding to mixed BADs is proven in a very similar but much easier manner[5].

Throughout this section, $n, d, \Delta, K$ denote integers satisfying: $0 < d < n$, $\Delta = \gcd(n, d)$ and $K = n/\Delta > 1$ ($n$ corresponds to the integer $\omega$ of the previous sections on BANs), and $(\Delta, K) \notin \{(1, 10), (2, 6)\}$.

$\mathbf{F}_1 = \{00, 111\}$ is easier to manipulate than $\mathbf{F}_d, d > 1$. For this reason, given a word $w \in \mathbf{W}_d^n$, the baseline idea is going to be to see $w$ as an interleaving $L$ of a certain number $m$ of smaller words of length $k$, $L(0), \ldots, L(m-1)$ such that $\forall 0 \le j < m$, letters of sub-word $L(j)$ appear every $d$ position in $w$, and $L(j) \in \mathbf{W}_1^k$ (*cf.* Fig. 2). As we are about to see, $m = \Delta$ and $k = K$.

---

[4] They always do if $(d, n), (n - d,) \notin \{(1, 10), (3, 10), (2, 12)\}$, *cf.* Footnote 3.

[5] The mixed BAD case involves less forbidden factors than the negative BAD case. Also, while the latter relates to the Perrin integer sequence which has a rugged beginning, the former relates to the much smoother Lucas sequence (*cf.* Table 1). For these reasons, the mixed BAD case induces no special sub-cases, especially none of the sort of $K = 6$ which is involved in the negative BAD case (*cf.* Appendix A.2 and A.6).

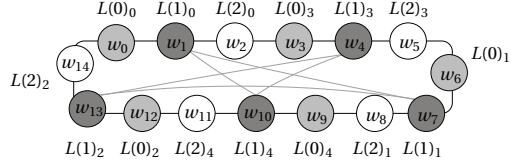

**Fig. 2.** A word $w \in \mathbb{B}^n$, $n = 15$, represented as an interleaving of $\Delta = \gcd(6, n) = 3$ words $L(1)$ (light grey), $L(2)$ (dark grey) and $L(3)$ (white) of length $K = 5$.

Generally, $\forall w \in \mathbb{B}^n$, we define the list $L(w, d) = \big(L(0), L(1), \ldots, L(\Delta-1)\big) \in (\mathbb{B}^K)^\Delta$ by: $\forall j < \Delta$, $\forall k < K$, $L(j)_k = w_{j+kd}$. And conversely, given a list $L = \big(L(0), \ldots, L(\Delta-1)\big) \in (\mathbb{B}^K)^\Delta$, we define the word $w(L, d)$ by: $\forall i < n$, $i = j + q\Delta \equiv j \bmod \Delta$, $w(L, d)_i = L(j)_{q \times \frac{\Delta}{d}}$. Note that by definition of $\Delta$ there exists Bezout integers $a, b \in \mathbb{Z}$ such that $\Delta = an + bd$. This implies that $\frac{\Delta}{d} \equiv b \bmod n$ so $q \times \frac{\Delta}{d} \in \mathbb{Z}/K\mathbb{Z}$. The following lemma can easily be checked.

**Lemma 2.** *For any word $w \in \mathbb{B}^n$ and any list $L \in (\mathbb{B}^K)^\Delta$, $w = w(L, d) \Leftrightarrow L = L(w, d)$.*

In the sequel, for any divisor $p$ of $n$, $\Delta_p$ and $K_p$ respectively denote the divisors of $\Delta$ and of $K$ defined by $\Delta_p = \gcd(p, d) = \gcd(\Delta, p)$ and $K_p = p/\Delta_p$, where $d = \ell \bmod p, d > 0$ (*cf.* Lemma 1) so that necessarily $K_p > 1$. The first part of the following result may be checked using the definitions of $\mathbf{F}_d$ and $\mathbf{F}_1$:

**Lemma 3.** *For any word $w \in \mathbb{B}^n$, $w \in \mathbf{W}_d^n \Leftrightarrow L(w, d) \in (\mathbf{W}_1^K)^\Delta$. Consequently:*

$$\mathbf{W}_d^n \simeq (\mathbf{W}_1^K)^\Delta \simeq \mathbf{W}_\Delta^n \quad \text{and} \quad \mathbf{W}_d^n(p) \simeq \mathbf{W}_d^p \simeq (\mathbf{W}_1^{K_p})^{\Delta_p} \simeq \mathbf{W}_{\Delta_p}^p. \tag{10}$$

Equation (10) allows to concentrate on the case where $\Delta = d$, w.l.g. The rest of the proof of Proposition 1 consists of the following steps, detailed in Appendix A:

1. Relate the period $p = K_p \Delta_p$ of a word $w \in \mathbf{W}_\Delta^n(p)$ to the interleaving $L = L(w, \Delta)$ representing it. More precisely, show that $K_p$ equals the least common multiple of the primitive periods of $L$'s sub-words $L(j)$ (*cf.* Lemma 4), and that $\forall j < \Delta$, $L(j) \equiv L(j + \Delta_p)$ (*cf.* Lemma 6).

2. Define a unique representative list $\dot{L}(\langle w \rangle, \Delta)$ for every conjugacy class $\langle w \rangle$. Introduce set $\widetilde{\mathbf{L}}^{K, \Delta} = \{\dot{L}(\langle w \rangle, \Delta), \langle w \rangle \in \widetilde{\mathbf{C}}_\Delta^{K\Delta}\}$. As follows, define a map $\Gamma : \bigcup_{p | n, p < n} \widetilde{\mathbf{L}}^{K_p, \Delta_p} \to \widetilde{\mathbf{L}}^{K, \Delta}$ (*cf.* A.3) from which $\gamma$ can be derived directly (*cf.* item 7 below).

3. Let $L \in \widetilde{\mathbf{L}}^{K_p, \Delta_p}$ ($p | n$, $p < n$) be an arbitrary list for which we must define an image by $\Gamma$. First, as follows, define $L' \in (\mathbf{W}_1^K)^\Delta$ so that $w = w(L', \Delta)$ is primitive.

(a) Define $L'$ in the case where $\Delta_p = \Delta$ (*cf.* A.4). To do this, first elongate one of the words $L(\mathbf{j}^*) \in \mathbf{W}_1^{K_p}$ of list $L$ so as to turn it into an primitive word $L'(\mathbf{j}^*) \in \widetilde{\mathbf{W}}_1^K$ (*cf.* the elongation map: $u \in \mathbf{W}_1^{K_p} \mapsto \alpha(K, u) \in \mathbf{W}_1^K$ defined in A.4). Except in (many) special cases, this elongation can be done by concatenating a primitive word of length $K - K_p$ to $L(\mathbf{j}^*)$. Next, repeat all other words to make them longer: $\forall j < \Delta_p,\; j \neq \mathbf{j}^*$, define $L'(j) = L(j)^{\frac{K}{K_p}}$.

(b) Define $L'$ in the case where $\Delta_p < \Delta$ (*cf.* A.5). To do this, first, if necessary (it could be that $K_p = K$ in this case), lengthen all $L(j)$'s by repeating them: $\forall j < \Delta_p,\; L'(j) = L(j)^{\frac{K}{K_p}}$. Then, add a series of $\Delta - \Delta_p \geq \Delta/2$ consecutive, identical primitive words $\mathbf{z} \in \widetilde{\mathbf{W}}_1^K$, in a way that this series will not be confused with the rest of $L'$.

4. In both cases 3a and 3b, primitivity of $w = w(L', \Delta)$ follows from item 1 above and from: *(i)* at least one of the $L'(j)$'s is primitive, and *(ii)* $L'$ is aperiodic itself by construction (in the general cases, a series of 1 periodic and $\Delta - 1$ aperiodic sub-words cannot be periodic, nor can a series in which at least half of the sub-words are identical, consecutive and distinct from the non-empty rest of the series).

5. Define $\Gamma(L) = \dot{L}(\langle w \rangle, \Delta)$ where $w = w(L', \Delta)$.

6. The injectivity of $\Gamma$ comes from the effort made in the construction of $L'$ to encode non-ambiguously all information of $L$ into $L'$[6], and from the fact that the domain of $\Gamma$ only contains one list $L = \dot{L}(\langle u \rangle, \Delta)$ per conjugacy class $\langle u \rangle$, $u \in \widetilde{\mathbf{W}}_{\Delta_p}^n$.

7. Define map $\gamma : \bigcup_{p|n, p<n} \widetilde{\mathbf{C}}_{\Delta_p}^p \to \widetilde{\mathbf{C}}_\Delta^n$ by $\gamma(\langle u \rangle) = \langle w \rangle$ where $w = w(\Gamma(\dot{L}(\langle u \rangle, \Delta_p)), \Delta)$, $\forall \langle u \rangle \in \widetilde{\mathbf{C}}_{\Delta_p}^p$ ($p|n,\; p < n$). The injectivity of $\gamma$ follows from that of $\Gamma$. □

## *12 Back to Boolean automata networks*

With Theorem 1 informing on the behaviours of BACs and BADs, it still remains to gain insight on the role played specifically by the cycle intersections in the defining of network (asymptotic) behavioural possibilities. To do this, comparisons between BACs and BADs need to be made. Our last formal result below (whose yet unpublished proof is given in Appendix B) exploits Theorem 1 to go further in this direction. We let $\mathtt{A}^+(\omega)$, $\mathtt{A}^-(\omega)$, $\mathtt{A}^{++}(\omega)$, $\mathtt{A}_\Delta^{-+}(\omega)$ and $\mathtt{A}_\Delta^{--}(\omega)$ respectively denote the total numbers of attractors of a positive BAC, of a negative BAC, of a positive BAD, of a mixed BAD and of a negative BAD of order $\omega$ (and such that $\Delta = \gcd(\ell, r)$).

**Theorem 2.** *The numbers of attractors of negative BACs (resp. of BADs) are (resp. exponentially w.r.t. $\omega$) smaller than that of positive BACs of same order $\omega \in \mathbb{N}$:*

$$\mathtt{A}^-(\omega) \leq \tfrac{1}{2}\mathtt{A}^+(\omega) \qquad \mathtt{A}_\Delta^{-+}(\omega) \leq 2\left(\tfrac{\sqrt{3}}{2}\right)^\omega \mathtt{A}^+(\omega)$$

$$\mathtt{A}^{++}(\omega) = \mathtt{A}^+(\omega) \qquad \mathtt{A}_\Delta^{--}(\omega) \leq 2\left(\tfrac{3^{\frac{1}{3}}}{2}\right)^\omega \mathtt{A}^+(\omega) \text{ if } K = 3$$

$$\mathtt{A}_\Delta^{--}(\omega) \leq 2\left(\tfrac{\sqrt{2}}{2}\right)^\omega \mathtt{A}^+(\omega) \text{ if } K \neq 3$$

---

[6] This effort and the many special cases that need to be taken into account separately are the explanation for the great length of the full proof given in Appendix A.

*where, for a* BAD, $\Delta$ *denotes the* gcd *of its underlying cycle lengths, and* $\omega = K\Delta$.

## 13 Discussion

*Informal insights and scope of results.*

By Theorem 1, *the largest attractors of a* BAC *or* BAD *are the most numerous*. Let $U(x) = \{i < n,\ x_i \neq f_i(x)\}$ denote the set of *local instabilities* in configuration $x$. Intuitively, *amongst the large attractors of a* BAC *or* BAD *must feature the most stable*, *i.e.* those involving configurations with small values of $\#U(x)$. The idea is that this sort of attractor, induced by little momentum, corresponds to a small number $\#U(x)$ of local instabilities, circulating on the cycles of **G**, punctually destabilising each automaton one after the other, before returning to their initial locations. This agrees with the fact that the order of a BAC or BAD *has the order* of its size (*cf.* Table 1).

Let us imagine turning BACs into BADs, for instance by forcing two automata to work as one, either adjoining two simple cycles, or pursing one large cycle into two smaller ones. This way, the order $\omega$ of the overall network, and, much more significantly, by Theorem 2, the total number $\text{A}(\omega)$ of attractors decrease. Backed up with simulation relations between BACs and BADs established in [18], Theorem 2 serves as as a base case supporting the following informal idea: *networks tend to lose degrees of freedom as their underlying structural cycles become more intricately intersected*. Put in other terms, this just means that cycles that are forced to interact tend to hinder themselves rather than the contrary, and Theorem 2 represents a first formal argument in this direction and in the context of BANs.

The intersection automaton 0 of a BAD can receive at most $2 = \deg^-(0)$ local instabilities as inputs from automata $\ell - 1$ and $n - 1$. It can output at most $2 = \deg^+(0)$. Examining the different cases reveals that, $\#U(x)$ is less often increased than it is maintained or decreased. It seems that all in all, BAD intersections tend to synchronise local instabilities and reduce the number of them.

Now, let us use the ratio $\xi(\mathcal{N}) = \frac{\widetilde{T}(\omega)}{\omega}$ to pinpoint formally a general notion of *degree of freedom* (or *propensity to behave in numerous, various ways*) of a BAN $\mathcal{N}$. As we turn BACs into BADs, the size $n$ of the overall network hardly changes at all. The order $\omega$ doesn't change much either since it still has the order of $n$. But on the contrary, by Theorem 2, $\xi(\mathcal{N})$ is very significantly decreased. We can build on all the previous remarks of this section, assuming that small attractors are induced by greater numbers of local instabilities, and that cycle intersections filter out both global and local instabilities. Thus, comparing BACs and BADs, the substantial difference in $\xi(\mathcal{N})$ that comes with no substantial change of $\omega$ can be interpreted as follows: *by getting rid of local instabilities, cycle intersections induce larger attractors and get rid of the smaller, less stable ones*. This would mean that even under the parallel updating which is the best at entertaining local instabilities on uninterrupted paths of **G**, cycle intersections "force" asynchrony in the sense that they reduce the number of possible changes that are possible at once (note that this is a sort of asynchrony that is *inherent* to the system rather than an assumption of the practitioner), thereby reducing the overall network asymptotic degree of freedom, and increasing its overall stability.

*Perspectives.*

First of all, of course, all the semantic remarks made right above call for a proper formalisation and a verification. One way or the other, I believe these remarks to be noteworthy since at the very least they can serve as very tangible (and new) guidelines for further researches. Moreover, one practical purpose can be expected to be drawn from the results of this paper, in the lines of these informal remarks: to yield a *constructive* method for approximating networks, based on elementary operations (including elementary operations on digraphs **G**, not dissimilar to the contractions underlying the definition of a graph minor) that "simplify" the networks (structures), and as a consequence, in a controlled manner, add noise in the description of their behaviours. Besides the pertinence of this with regards to modelling considerations where complexity is especially limiting, this would allow to derive bounds on the numbers of attractors of arbitrary networks.

As for the technical aspects of this paper, we have built an injective map from a set of periodic binary necklaces satisfying certain conditions to the set of primitive binary necklaces satisfying the same conditions. This raises the problem of specifying more generally what are the types of conditions on necklaces that allow this to remain true (just like it was already known to be true for necklaces satisfying no conditions at all [24]). Perhaps tightly related to this but with a different viewpoint is the following problem which arises from Lemma 7 and Equation (21) in Apprendix A. All five cases considered in this paper and shown to fall under the scope of Theorem 1 are based on integer sequences $(X(n))_{n \in \mathbb{N}}$ whose value in $n \in \mathbb{N}$ either equals or is very close and asymptotically equivalent to the $n^{th}$ power $a^n$ of some value $a \in \mathbb{R}$, $1 < a$. We have found no mention in the literature of a more general result characterising a larger class of integer sequences $(X(n))_{n \in \mathbb{N}}$ satisfying Theorem 1. This is one the most immediate sequels to the present work.

# Appendix A

## A  Proof of Proposition 1

### A.1  Lists and words (sequel)

We let integers $n, \Delta, K, p, \Delta_p, K_p > 1$ be as before. In particular $(\Delta, K) \notin \{(1, 10), (2, 6)\}$.

Lemmas 4 and 6 show how the (primitive) period $p$ of a word $w \in \mathbf{W}_\Delta^n(p)$ translates in terms of the interleaving $L = L(w, \Delta)$. First, Lemma 4 relates $p$ to the primitive period of the interleaved words $L(j)$:

**Lemma 4.**  *If $p = K_p \Delta_p$ ($\Delta_p = \gcd(\Delta, p)$) is the primitive period of word $w \in \widetilde{\mathbf{W}}_\Delta^n(p)$, then $K_p = \mathrm{lcm}_{j<\Delta}(K_j)$ where, $\forall j < \Delta$, $K_j$ denotes the primitive period of word $L(j) \in \mathbf{W}_1^K$ of list $L(w, \Delta)$.*

*Proof.*  Let $K' = \mathrm{lcm}_{j<\Delta}(K_j)$. On the one hand, $\forall j < \Delta, \forall k < K$, $L(j)_{k+K_p} = w_{j+(k+K_p)\Delta} = w_{j+k\Delta + p\frac{\Delta}{\Delta_p}} = w_{j+k\Delta} = L(j)_k$. Thus, $K_p$ is a common period of all $L(j)$'s and $K' | K_p$. On the other hand, $\forall i < n, i = j + q\Delta \equiv j \bmod \Delta$, $w_{i+K'\Delta} = L(j)_{q+K'} = L(j)_q = w_i$ so $K'\Delta$ is a period of $w$ and $p | K'\Delta$ which implies $K_p | K$ because $\gcd(K_p, \Delta/\Delta_p) = 1$. $\square$

Further, Lemma 6 relates $p$ to the *interleaving* of words $L(j)$. To do that, it exploits Lemma 5 which relates the list $L^q = L(\rho^q(w), \Delta)$ associated to an arbitrary conjugate $\rho^q(w)$ of a word $w \in \mathbb{B}^n$, with the list $L = L(w, \Delta)$:

**Lemma 5.**  *The words in the list representing an arbitrary conjugate $w' = \rho^q(w)$ of $w$ are rotations of the words in the list representing $w$, i.e. $w'$ and $w$ are interleavings (w.r.t. $\Delta$) of rotations of the same sub-words: if $q = m\Delta + \delta \equiv \delta \bmod \Delta$, then:*

$$L^q = L(w', \Delta) = \big( \underbrace{\rho^{m+1}(L(\Delta-\delta)), \ldots, \rho^{m+1}(L(\Delta-1))}_{\delta \text{ words } L^q(j) = \rho^{m+1}(L(j+\Delta-\delta)),\ 0 \le j < \delta}, \overbrace{\rho^m(L(0)), \ldots, \rho^m(L(\Delta-\delta-1))}^{\Delta-\delta \text{ words } L^q(j) = \rho^m(L(j-\delta)),\ \delta \le j < \Delta} \big).$$

*Proof.*  If $j < \delta$, then $0 < \Delta + j - \delta < \Delta$ so $\forall k < K$, $L^q(j)_k = \rho^q(w)_{j+k\Delta} = w_{j+k\Delta-q} = w_{(j-\delta+\Delta)+(k-1-m)\Delta} = L(j-\delta+\Delta)_{k-(m+1)} = \rho^{m+1}(L(j-\delta+\Delta))_k$. If $\delta \le j < \Delta$, then $0 < j - \delta < \Delta$ so $\forall k < K$, $L^q(j)_k = L(j-\delta)_{k-m} = \rho^m(L(j-\delta))$. $\square$

**Lemma 6.**  $\forall w \in \mathbf{W}_\Delta^n(p)$, $L(j + \Delta_p) \equiv L(j)$. *More precisely, $L(j + \Delta_p) = \rho^{\frac{\Delta_p}{\Delta}}(L(j))$ if $j < \Delta - \Delta_p$ and $L(j + \Delta_p) = \rho^{\frac{\Delta_p}{\Delta}-1}(L(j))$ otherwise.*

*Proof.*  By definition, there exists Bezout integers $a, b \in \mathbb{Z}$ s.t. $\Delta_p = a\Delta + bp$. Lemma 6 is proven by using Lemma 5, noting that $w \in \mathbf{W}_\Delta^{n,p} \Leftrightarrow \rho^{kp}(w) = w$, $\forall k \in \mathbb{N}$, and taking $k = b$. $\square$

## A.2 Primitive words

In this section we look closer at some primitive words of $\widetilde{\mathbf{W}}_a^b$, depending on $a$ and $b$. First let us recall that $\forall a, \widetilde{\mathbf{W}}_a^a = \mathbf{W}_a^a = \emptyset$. Next, let us consider the cases where $K$ equals either 4 or 6.

Since $\mathbf{W}_1^4 = \langle 0101 \rangle = \mathbf{W}_1^4(2)$, it holds that $\widetilde{\mathbf{W}}_1^4 = \emptyset$, and, by Lemma 3 and Equation (7), that $\mathbf{W}_\Delta^{4\Delta} \sim (\mathbf{W}_1^4)^\Delta \sim (\mathbf{W}_1^4(2))^\Delta \sim (\mathbf{W}_1^2)^\Delta \sim \mathbf{W}_\Delta^{2\Delta} \sim \mathbf{W}_\Delta^{4\Delta}(2\Delta)$ so there are no primitive words in $\widetilde{\mathbf{W}}_\Delta^n$ when $n = 4\Delta$: $\forall \Delta, \widetilde{\mathbf{W}}_\Delta^{4\Delta} = \emptyset$.

Since $\mathbf{W}_1^6 = \langle 010101 \rangle \cup \langle 011011 \rangle = \widetilde{\mathbf{W}}_1^6(2) \cup \widetilde{\mathbf{W}}_1^6(3)$, it holds that $\forall \Delta, \widetilde{\mathbf{W}}_1^6 = \emptyset$. And if $w \in \widetilde{\mathbf{W}}_\Delta^{6\Delta}$, then $L(w, \Delta)$ is composed of words of period 2 or 3 and contains at least one of each.

We recall that $K > 1$ necessarily holds because $\omega = n = K\Delta$ cannot divide $\ell$ and thus nor can it divide $\Delta$ (*cf.* Lemma 1). From now on, $K \notin \{1, 4\}$ is assumed and the case $K = 6$ will be treated separately. $\forall K \notin \{1, 4, 6\}$, let us define the following four canonical words of $\mathbf{W}_1^K$ that can be checked to be primitive:

$$\begin{aligned}
\mathbf{u}(K) &= (01)^a (011)^b \text{ where } a \text{ is maximal so that } K = 2a + 3b, \text{ with } b \leq 2 \\
\mathbf{x}(K) &= \mathbf{u}(K-5)\mathbf{u}(5), \quad \forall K \geq 12, \\
\mathbf{v}(K) &= (01)^a (011)^b \text{ where } b \text{ is maximal such that } K = 2a + 3b, \text{ with } a \leq 3 \\
\mathbf{y}(K) &= \mathbf{v}(K-7)\mathbf{v}(7), \quad \forall K > 14.
\end{aligned} \quad (11)$$

Let us call **macro-letter** any of the two factors $01$ and $011$. Any word of $\mathbf{W}_1^K$ is a word on alphabet $\{01, 011\}$.

Comparing $a$ and $b$ in the writing of $\mathbf{u}$ and $\mathbf{v}$ and also, comparing the number of alternations of macro-letters, the following may be proven easily and will serve as a basis to the map $\Gamma$ on lists built in the next paragraphs:

$$\begin{aligned}
\forall K \leq 10, \quad \widetilde{\mathbf{W}}_1^K &= \langle \mathbf{u}(K) \rangle = \langle \mathbf{v}(K) \rangle \\
\widetilde{\mathbf{W}}_1^{11} &= \langle \mathbf{u}(11) \rangle \cup \langle \mathbf{v}(11) \rangle && \text{where } \mathbf{u}(11) \not\equiv \mathbf{v}(11) \\
\widetilde{\mathbf{W}}_1^{12} &= \langle \mathbf{u}(12) \rangle \cup \langle \mathbf{x}(12) \rangle && \text{where } \mathbf{u}(12) \equiv \mathbf{v}(12) \not\equiv \mathbf{x}(12) \\
\forall K > 12, \quad \widetilde{\mathbf{W}}_1^K &\supset \langle \mathbf{u}(K) \rangle \cup \langle \mathbf{v}(K) \rangle \cup \langle \mathbf{x}(K) \rangle && \text{and } \mathbf{u}(K) \not\equiv \mathbf{v}(K) \not\equiv \mathbf{x}(K) \not\equiv \mathbf{u}(K) \\
\forall K > 14, \quad \widetilde{\mathbf{W}}_1^K &\supset \langle \mathbf{u}(K) \rangle \cup \langle \mathbf{v}(K) \rangle \cup \langle \mathbf{x}(K) \rangle \cup \langle \mathbf{y}(K) \rangle && \text{and } \langle \mathbf{y} \rangle \notin \{\langle \mathbf{u}(K) \rangle, \langle \mathbf{v}(K) \rangle, \langle \mathbf{x}(K) \rangle\}
\end{aligned}$$

$$\forall u \in \mathbf{W}_1^{K_p}, K_p \notin \{2, 3\}, \quad \langle \mathbf{u}(K - K_p) u \rangle \notin \{\langle \mathbf{u}(K) \rangle, \langle \mathbf{v}(K) \rangle, \langle \mathbf{x}(K) \rangle\}. \quad (12)$$

## A.3 The maps $\Gamma$ and $\Gamma'$ on lists

If we were to map primitive words $u \in \widetilde{\mathbf{W}}_\Delta^p \subset \mathbb{B}^p$, with $p < n$, injectively onto primitive words $w \in \widetilde{\mathbf{W}}_\Delta^n \subset \mathbb{B}^n$, we could do this through the construction of a one-to-one map belonging to $\bigcup_{p|n, p<n} (\mathbf{W}^{K_p})^{\Delta_p} \to (\mathbf{W}^K)^\Delta$ on lists representing primitive words. Here, to prove (9), we just want to map primitive *necklaces* $\langle u \rangle \in \widetilde{\mathbf{C}}_\Delta^p$ injectively onto primitive necklaces $\langle w \rangle \in \widetilde{\mathbf{C}}_\Delta^n$. To do this, we are going to build a map on lists that represent primitive necklaces. Thus, we must define the representative list $\dot{L}(\langle w \rangle, \Delta)$ of an arbitrary necklace $\langle w \rangle$.

For that purpose, we order words and list of words in $\mathbf{W}_1^K$ lexicographically (*i.e.* w.r.t. order $\prec$ on letters defined by $0 \prec 1$, or w.r.t. to order $\prec'$ on macro-letters defined by $01 \prec' 011$). For any $u \in \mathbf{W}_1^K$, let us denote by $\dot{u}$ the smallest word in $\langle u \rangle$.

We define $\dot{L}(\langle w \rangle, \Delta)$ to be the smallest of the lists in $\{L(w', \Delta), \ w' \equiv w\}$ such that:

$$L(0) = \dot{L}(0) \text{ and } \forall j < \delta, \ L(0) \leq \dot{L}(j) \tag{13}$$

(where $\leq$ is the lexicographical order on words induced by $\prec$). It follows from Lemma 5 that such a list exists indeed. Then, we can naturally introduce the following sets of lists:

$$\mathbf{L}^{K,\Delta} = \{\dot{L}(\langle w \rangle, \Delta), \ \langle w \rangle \in \mathbf{C}_\Delta^{K\Delta}\} \quad \text{and} \quad \widetilde{\mathbf{L}}^{K,\Delta} = \{\dot{L}(\langle w \rangle, \Delta), \ \langle w \rangle \in \widetilde{\mathbf{C}}_\Delta^{K\Delta}\}.$$

And now we aim at defining a map $\Gamma : \bigcup_{p|n, p<n} \widetilde{\mathbf{L}}^{K_p, \Delta_p} \to \widetilde{\mathbf{L}}^{K,\Delta}$. To do this, we first want to define a map $\Gamma' : \bigcup_{p|n, p<n} \widetilde{\mathbf{L}}^{K_p, \Delta_p} \to L(\widetilde{\mathbf{W}}_\Delta^n, \Delta)$, with images in the set of (non necessarily representative) lists associated to primitive words of $\widetilde{\mathbf{W}}_\Delta^n$. From the definition of this map $\Gamma'$, the definition of $\Gamma$ will immediately be given by:

$$\Gamma(L) = \dot{L}(\langle w \rangle, \Delta) \text{ where } w = w(\Gamma'(L), \Delta). \tag{14}$$

We thus need to turn a list $L$ of $\Delta_p$ words of length $K_p$ into a list $L'$ of $\Delta$ words of length $K$ (and take the representative $\dot{L}'$ of the latter).

Notably, this list $L$ we want to build images $L'$ and $\dot{L}'$ by $\Gamma'$ and $\Gamma$, need not be any list: it must satisfy (13) and be a *representative list* $L = \dot{L}(\langle u \rangle, K_p)$, representing a *primitive* necklace $\langle u \rangle$ of length $p$, where, importantly, $p | n, \ p < n$ is a proper divisor of the length $n$ of the necklace $\langle w \rangle = \langle w(L', \Delta) \rangle = \langle w(\dot{L}', \Delta) \rangle$. And we recall that as before, $p$ does not divide $\Delta$, implying that $K_p > 1$.

To build $L'$ from $L$, we want to add $K - K_p$ letters to the sub-words $L(j) \in \mathbf{W}_1^{K_p}$ of $L$, and add $\Delta - \Delta_p$ new words. In the general case, $L(j)$ will just be repeated to create $L'(j) = L(j)^{K/K_p} \in \mathbf{W}_1^K(K_p)$. As for the $\Delta - \Delta_p$ added words, we will choose them to be primitive in order to ensure the primitivity of the resulting word $w = w(L', \Delta)$ (*cf.* Lemma 4). Of course this only works when $\Delta_p < \Delta$. When $\Delta_p = \Delta$, by Lemma 4, words in $L'$ cannot keep on having common period $K_p$, otherwise $w$ would have period $K_p \Delta < n$. So in this case, we still must ensure the primitivity of $w$, but we must do it without adding any words. For this reason, when $\Delta_p = \Delta$ (implying $K_p < K$) we are going to *elongate* exactly one of the $L(j)$'s into a primitive word $\alpha(K, L(j)) = L'(j) \in \widetilde{\mathbf{W}}_1^K$. And we are going to do it in a way that, given only $L'(j)$, it is still possible to derive what $L(j)$ was $L'(j)$ constructed from (to ensure injectivity of $\Gamma$). So we want there to exist a map $\beta : \alpha(K, L(j)) \in \mathbf{W}_1^K \mapsto L(j) \in \bigcup_{K_p | K} \mathbf{W}_1^{K_p}$. The baseline idea of this *elongation* is to concatenate word $\mathbf{u}(K - K_p)$ or word $\mathbf{v}(K - K_p)$ – *cf.* (11) – to word $L(j)$ so that the elongated version of $L(j)$ looks like: $L'(j) = \alpha(K, L(j)) \equiv (01)^a (011)^b L(j)$ and is primitive. However, the injectivity of $\Gamma$ and primitivity of $w(L', \Delta)$ require that this idea be adjusted carefully in some cases.

### A.4 Case $\Delta_p = \Delta$

In this case, $K_p | K > K_p$ so in particular, $K$ is not prime. Also, we assume that $K \notin \{1, 4, 6\}$. We want to elongate a word $u = L(j) \in \mathbf{W}_1^{K_p}$ of length $K_p$ into a primitive word $\alpha(K, u) = L'(j) \in \widetilde{\mathbf{W}}_1^K$ of length $K$. Equation (12) limits the choice for the latter: $\forall K \leq 10$, there is only one primitive word of length $K$, and for $K = 12$, there only are two. However, as a result of assumptions, we are only considering the following cases
(otherwise, $K$ is prime or belongs to $\{1,4,6\}$):

- $K = 8$ implying that $K_p = 2$ (by A.2),
- $K = 9$ and $K_p = 3$,
- $K = 10$ and $K_p \in \{2,5\}$,
- $K = 12$ and $K_p \in \{2,3,6\}$,
- $K = 14$ and $K_p \in \{2,7\}$, and
- $K > 14$.

Thus, $\forall u \in \mathbf{W}_1^{K_p}$, we define $\alpha(K,u) \in \mathbf{W}_1^K$ as follows.

When $K_p \notin \{2,3,6\}$:
$$\forall \dot{u} \in \mathbf{W}_1^{K_p}, \alpha(K,\dot{u}) = \mathbf{u}(K-K_p)\dot{u} \quad \text{if } \dot{u} \neq \mathbf{u}(\tfrac{K}{2})$$
$$\alpha(K,\mathbf{u}(\tfrac{K}{2})) = \mathbf{v}(K) \quad \text{(noting that for } \mathbf{u}(\tfrac{K}{2}) \text{ to exist, it must hold that}$$
$$K \neq 12, \text{ i.e. } K = 10 \vee K > 12)$$

When $K_p \in \{2,3,6\}$:
$$\alpha(K,01) = \alpha(K,010101) = \mathbf{u}(K)$$
$$\alpha(K,011) = \alpha(K,011011) = \begin{cases} \mathbf{u}(K) & \text{if } K < 12 \\ & \text{and } 3|K, \text{ i.e. } K = 9 \\ \mathbf{x}(K) & \text{if } K \geq 12 \\ & \text{and } 3|K, \text{ i.e. } K = 12 \vee K > 14 \end{cases}$$

Generally, $\forall k \in \mathbb{Z}/K_p\mathbb{Z}, \forall u = \rho^k(\dot{u}) \in \mathbf{W}_1^{K_p}$, we let: $\alpha(K,u) = \rho^k\big(\alpha(K,\dot{u})\big)$.

It follows from the definitions of $\mathbf{u}(k), \mathbf{v}(k)$ and $\mathbf{x}(k)$ in (11) and from (12) that $\alpha(K,u)$ is primitive $\forall K \notin \{1,4,6\}, \forall u \in \mathbf{W}_1^{K_p}$.

Let $K_j$ denote again the primitive period of word $L(j)$ of $L$, and let $\overline{K}_j = \text{lcm}\{K_i, i \neq j\}$. If $K_p \notin \{2,3,6\}$, let $\mathbf{j}^* < \Delta = \Delta_p$ be such that $K_{\mathbf{j}^*} \notin \{2,3\}$. Otherwise let $\mathbf{j}^* = 0$. We define $L' = \Gamma'(L)$ by:
$$\begin{cases} L'(\mathbf{j}^*) = \alpha(K, L(\mathbf{j}^*)) \\ L'(j) = L(j)^{\frac{K}{K_p}}, \quad \forall j \neq \mathbf{j}^* . \end{cases} \quad (15)$$

Now, as follows we define $\beta$ (for the general case where $K \neq 10$) and $\beta'$ (for case $K = 10$) to retrieve $L(j)$ from $L'(j) = \alpha(K, L(j))$:

$$\begin{array}{ll}
\forall K \geq 12, \ \beta(\mathbf{u}(8)) = \beta(\mathbf{u}(K)) = 01 & \beta'(2, \mathbf{u}(10)) = \quad 01 \\
\quad \beta(\mathbf{u}(9)) = \beta(\mathbf{x}(K)) = 011 & \beta'(5, \mathbf{u}(10)) = \quad 01011 \\
\forall K > 12, \ \beta(\mathbf{v}(K)) = \mathbf{u}(\tfrac{K}{2}) & \text{and } \forall w = \rho^q(\dot{w}), \\
\forall K > 12, \ \forall \dot{u}, \ \beta(\mathbf{u}(K')\dot{u}) = \dot{u} & \beta'(k,w) = \rho^q(\beta(k,\dot{w})). \\
\text{and } \forall w = \rho^q(\dot{w}), \ \beta(w) = \rho^q(\beta(\dot{w})) &
\end{array} \quad (16)$$

Unless $u \in \mathbf{W}_1^{K_p}$ has period 2 or 3, $\beta(\alpha(K,u)) = u$, and $\beta'(|u|, \alpha(10,u)) = u$.

Examining $L' = \Gamma(L)$, $K_p$ can be derived in each case as follows (which will be useful in A.8:

- $K = 10$ $\implies K_p \in \{2,5\}$ and $K_p = K_j$, $\forall j < \Delta$
- $\overline{K}_{\mathbf{j}^*} \notin \{2,3,6\}$ $\implies K_{\mathbf{j}^*} \notin \{2,3\}$ $\implies K_p = \text{lcm}\{K_{\mathbf{j}^*}, \overline{K}_{\mathbf{j}^*}\} \notin \{2,3,6\}$
- $\overline{K}_{\mathbf{j}^*} = 6$ $\implies K_p = 6$
- $\overline{K}_{\mathbf{j}^*} \in \{2,3\}$ $\implies$ either $\overline{K}_{\mathbf{j}^*} \neq K_{\mathbf{j}^*}$ and thus $K_p = \text{lcm}\{K_{\mathbf{j}^*}, \overline{K}_{\mathbf{j}^*}\} = 6$ ($\implies K \geq 12$)
  or $\overline{K}_{\mathbf{j}^*} = K_{\mathbf{j}^*}$ and thus $\forall j < \Delta$, $K_p = K_j \in \{2,3\}$

(17)

### A.5  Case $\Delta_p < \Delta$

Here, we will not elongate any word of $L$. In the main case, the idea is to insert a series of $\Delta - \Delta_p$ new identical primitive words $\mathbf{z} \in \widetilde{\mathbf{W}}_1^K$ at a certain index $\mathbf{j}^* < \Delta_p$ of the list that will guarantee the primitivity of $w(L', \Delta)$. Note that unless $\Delta \leq 2$, we will be adding more than 1 primitive word. This way, $L'$ constructed in this case ($\Delta_p < \Delta$) cannot be confused with a $L'$ constructed in the previous ($\Delta_p = \Delta$) which only contains 1 primitive word (*cf.* A.4). The only possible ambiguity is when $\Delta = 2$ and $L'$ contains one primitive sub-word and one imprimitive sub-word. The resolution of this ambiguity will come later.

In the general case, we want $\mathbf{z}$ and $\mathbf{j}^*$ to be such that the range of added $\mathbf{z}$'s is distinguishable from the rest of $L'$ made from words of $L$. Given such a $\mathbf{z}$ and such a $\mathbf{j}^*$, we define $L' = \Gamma'(L)$ by:

$$\begin{cases} L'(j) = L(j)^{\frac{K}{K_p}} & \forall j < \mathbf{j}^* \\ L'(j) = \mathbf{z} & \forall \mathbf{j}^* \leq j < \mathbf{j}^* + \Delta - \Delta_p \\ L'(j) = L(j - \Delta + \Delta_p)^{\frac{K}{K_p}} & \forall j \geq \mathbf{j}^* + \Delta - \Delta_p. \end{cases} \quad (18)$$

When $K > 14$, we let $\mathbf{j}^* = 1$ and $\mathbf{z} = \rho^q(\mathbf{y}(\mathbf{K}))$ (word $\mathbf{y}(\mathbf{K})$ is defined in (11)) where $q$ is such that $\mathbf{z} \notin \{L'(0), L'(1)\}$. This is always possible because since $\mathbf{y}(\mathbf{K})$ is primitive, it holds that $|\langle \mathbf{y}(\mathbf{K})\rangle| = K > 14 > 2 = |\{L'(0), L'(1)\}|$.

If $K \leq 14$ and $\forall K \notin \{1,4,6\}$, we let $\mathbf{j}^*$ and $\mathbf{z}$ be such that (*cf.* Lemma 5):

$$\begin{cases} \mathbf{j}^* > 0 \text{ and } \mathbf{z} \notin \{L(\mathbf{j}^* - 1), L(\mathbf{j}^*)\} & \text{or} \\ \mathbf{j}^* = 0 \text{ and } \mathbf{z} \notin \{\rho(L(\Delta_p - 1)), L(0)\}. \end{cases} \quad (19)$$

This is always possible. Indeed, if not, *i.e.* if there is no such $\mathbf{j}^*$ and $\mathbf{z}$, then $K_p = K$ must hold since $\mathbf{z}$ has length $K$ and the $L(j)$'s have length $K_p$. Also, we must have $\forall 0 < j < \Delta_p$, $\widetilde{\mathbf{W}}_1^K = \{L(j), L(j+1)\} = \{L(0), \rho(L(\Delta_p - 1))\}$, and thus $|\mathbf{W}_1^\mathbf{K}| \leq 2$. As a result, $K = K_p = 2$ holds and $\Delta_p$ has to be odd (so that $L(0) = L(\Delta_p - 1) \neq \rho(L(\Delta_p - 1))$). But in this case, we can show that:

$$u = w(L, \Delta_p) \equiv \underbrace{(01)^{\frac{\Delta_p - 1}{2}} 0}_{\substack{1^{st} \text{ letters of words} \\ L(j),\ 0 \leq j < \Delta_p}} \underbrace{(10)^{\frac{\Delta_p - 1}{2}} 1}_{\substack{2^{nd} \text{ and last letters} \\ \text{of words } L(j),\ 0 \leq j < \Delta_p}} \equiv (01)^{\Delta_p} \notin \widetilde{\mathbf{W}}_{\Delta_p}^{2\Delta_p}$$

has period 2 rather than 2Δ. It is imprimitive and like all imprimitive words, its associated list $L$ needs no image by Γ. Thus, when $K \leq 14$, there always are $\mathbf{j}^*$ and $\mathbf{z}$ satisfying (19).

Let $w = w(L', \Delta)$. By Lemma 5, if $q = m\Delta + \Delta - \mathbf{j}^* \equiv \Delta - \mathbf{j}^* \mod \Delta$, then $w' = \rho^q(w)$ is characterised by a list *starting* with a series of at least $\frac{\Delta}{2}$ identical primitive words:

$$L^q = \Big( \underbrace{\ldots\ldots \rho^{m+1}(\mathbf{z}) \ldots\ldots}_{\Delta - \Delta_p \geq \frac{\Delta}{2} \text{ identical words}}, \underbrace{\ldots \rho^{m+1}(L(j)^{\frac{K}{K_p}})\ldots}_{\mathbf{j}^* \leq j < \Delta_p}, \underbrace{\ldots \rho^m(L(j)^{\frac{K}{K_p}})\ldots}_{0 \leq j < \mathbf{j}^*} \Big).$$

$L^{q+1}$, however, does not start with such a series. Indeed, $L^{q+1}(0) = \rho(L^q(\Delta - 1)) = \rho^{m+1}(L(\mathbf{j}^*)^{K/K_p}) \neq L^{q+1}(1) = \rho^{m+1}(\mathbf{z})$. Moreover, in $L^q$ (as in any $L^{q'}$) this longest series of identical words remains well bounded from the right since $L^q(\Delta - \Delta_p - 1) = \rho^{m+1}(\mathbf{z}) \neq L^q(\Delta - \Delta_p) = \rho^{m+1}(L(\mathbf{j}^*)^{\frac{K}{K_p}})$. As a result of this and of its long length, this series can be identified non-ambiguously in the list $L(v, \Delta)$ of any conjugate $v$ of $w$.

### A.6 Case $K = 6$

Importantly, in this case, $\Delta > 1$ necessarily holds. In addition, we assume that $\Delta \neq 2$ so that $\Delta > 2$. This case is strongly inspired from the general case (A.4 and A.5). The main difficulty lies in that since $\widetilde{\mathbf{W}}_1^6 = \emptyset$, $L' = \Gamma'(L) \in (\mathbf{W}^6)^\Delta$ must be composed of *imprimitive* words of period 2 and 3.

Let $\mathbf{u}(6) = 010101$ and $\mathbf{v}(6) = 011011$ so that $\mathbf{W}_1^6 = \langle \mathbf{u}(6) \rangle \cup \langle \mathbf{v}(6) \rangle$.

1. If $\Delta_p = \Delta > 2$ (and $K = 6 > K_p \in \{2, 3\}$), rather than elongating a word of $L$ into a primitive word as before, we replace one of period $K_p$ with one of period $\frac{6}{K_p}$: we define $L' = \Gamma'(L)$ by:

   $$\begin{cases} L'(0) & = \mathbf{v}(6) \text{ if } K_p = 2 \text{ and } L(0) = 01 \\ L'(0) & = \mathbf{u}(6) \text{ if } K_p = 3 \text{ and } L(0) = 011 \\ L'(j) & = L(j)^{\frac{K}{K_p}}, \quad \forall j > 0. \end{cases} \tag{20}$$

   Thus, $L'$ contains $\Delta - 1$ sub-words of period 2 (or 3) and one sub-word of period 3 (resp. 2).

2. Otherwise, if $\Delta_p > 1$ and $D = \Delta - \Delta_p > 2$, we add a series of $D$, identical words of period $\mathbf{z}$ of period $K_\mathbf{z} \in \{2, 3\}$. To do so, we let $\mathbf{j}^*$ and $\mathbf{z} \in \mathbf{W}_1^6$ be such that (19) holds. This is trivially possible if $K_p = 2$ (resp. 3) because we can just take $\mathbf{z} = \mathbf{v}(6)$ (resp. $\mathbf{z} = \mathbf{u}(6)$). Otherwise is also possible because $|\widetilde{\mathbf{W}}_1^6| > 2$. We define $L' = \Gamma'(L)$ as in (18). Let us note that if $\Delta_p = \Delta/2$, then $K_p \in \{3, 6\}$. Indeed, if $K_p = 2$, then $p = \Delta|\Delta$ which is impossible. As a consequence, only one list $L$ is mapped onto list $L' = (\mathbf{v}(6)^{\Delta/2}, \mathbf{u}(6)^{\Delta/2})$, that is list $L = (\mathbf{v}(6)^{\Delta/2})$.

   Since $\Delta_p > 1$, the series of words that we add has length $D < \Delta - 1$. The resulting list $L'$ cannot be confused with a list defined as in the previous case. Also, since we add more than two words, this case cannot be confused with the next ones either.

3. Otherwise, if $D = \Delta - \Delta_p \le 2$, then either $\Delta = 4$ and $\Delta_p = D = 2$, or $\Delta = 3$, $D = 2$ and $\Delta_p = 1$.

   In the fist case, $K_p \in \{3, 6\}$ necessarily holds (otherwise $p|\Delta$). If $K_p = 3$ holds, then $L = \big(011, 011\big) = L(001111, 2)$. In this case, we define $L' = \Gamma'(L) = \big(\mathbf{u(6)}, \mathbf{u(6)}, \mathbf{u(6)}, \mathbf{v(6)}\big) = L(0^4 1^4 0^3 1^4 0^4 1^5, 4)$. If $K_p = 6$ holds, then $L = \big(\mathbf{u}(6), \mathbf{v}(6)\big) = L(001101100111, 2)$. In this case, we define $L' = \Gamma'(L) = \big(\mathbf{v(6)}, \mathbf{v(6)}, \mathbf{v(6)}, \mathbf{u(6)}\big) = L(0^4 1^3 1^4 0^4 1^4 0 1^4, 4)$.

   In the second case, $K_p = 2$ (otherwise $p|\Delta$), and $L = \big(01\big)$. We define (consistently with the next item) $L' = \Gamma'(L) = \big(\mathbf{u(6)}, \mathbf{u(6)}, \mathbf{v(6)}\big) = L(0^3 1^3 0^2 1^3 0^3 1^4, 3)$.

4. In the remaining cases, $\Delta/2 > \Delta_p = 1$ and $K_p \in \{2, 3\}$. If $K_p = 2$ and $L = \big(01\big)$, then $\Delta$ must be odd (otherwise $p|\Delta$). In this case we define $L' = \Gamma'(L)$ by:

$$L' = \Big(\mathbf{u(6)}, \ \big(\mathbf{u(6)}, \mathbf{v(6)}\big)^{\frac{\Delta-1}{2}}\Big).$$

   If $K_p = 3$ and $L = \big(011\big)$, then $\Delta \bmod 3 = a \ne 0$ (otherwise $p|\Delta$). In this case we define $L' = \Gamma'(L)$ by:

$$L' = \Big(\mathbf{v(6)}^a, \ \big(\mathbf{v(6)}, \mathbf{u(6)}, \mathbf{v(6)}\big)^{\frac{\Delta-a}{3}}\Big)$$

   (where in particular, $L' = \big(\mathbf{v(6)}, \mathbf{v(6)}, \mathbf{u(6)}, \mathbf{u(6)}\big) = L(0^4 1^6 0^4 1^4 0^2 1^4, 4)$ if $\Delta = 4$). In both cases, $L'$ has at most two consecutive sub-words of same period.

### A.7 Primitivity of word $w(\Gamma(L), \Delta)$

Let $w = w(L', \Delta)$ and let $q = K_q \Delta_q$ be the primitive period of $w$. In all cases, the primitivity of at least one of the $L'(j)$'s guarantees that $K_q = K$ by Lemma 4. The *unicity* of the primitive word in $L'$ when $\Delta_p = \Delta$, and the long length ($\ge \Delta/2$) of the added well bounded series of consecutive identical primitive words when $\Delta_p < \Delta$, guarantee by Lemma 6 that $\Delta_q = \gcd(q, \Delta) = \Delta$. Thus, the primitivity of $w$ is ensured in all cases.

### A.8 Injectivity of $\Gamma$

Let $\langle w \rangle \in \widetilde{\mathbf{C}}^n_\Delta$ be a primitive necklace of length $n$. Let $\dot{L}' = L(\langle w \rangle, \Delta) \in \widetilde{\mathbf{L}}^{K,\Delta}$ be its representative list. We assume that $K \notin \{1, 4\}$. Algorithm 1 shows that there is at most one divisor $p = K_p \Delta_p < n$ of $n$ and one necklace $\langle u \rangle \in \mathbf{C}^p_{\Delta_p}$ such that $\dot{L}' = \Gamma(\dot{L}(\langle u \rangle, \Delta_p)) \in \Gamma(\bigcup_{p|n, p<n} \widetilde{\mathbf{L}}^{K_p, \Delta_p})$. □

**if** $K \neq 6$ **then**
    **if** *the number $\pi$ of primitive words $\dot{L}'(j)$ in $\dot{L}'$ equals $\pi = 1$* **then**
        Let $\mathbf{j}^*$ be the index of the only primitive word $\dot{L}'(\mathbf{j}^*) \in \widetilde{\mathbf{W}}_1^K$ of $\dot{L}'$.
        **if** *the number $\iota$ of imprimitive words $\dot{L}'(j)$ in $\dot{L}'$ equals $\iota = 0$* **then**
            $\Delta_p = 1 = \Delta$ and thus $\dot{L}'(0) = \alpha(K_p, u)$ for some $K_p < K$, $K_p \neq 6$, and some
            $u = L(0) \in \widetilde{\mathbf{W}}_1^{K_p}$. Recalling that $(\Delta, K) \neq (1, 10)$ is assumed, $L(0)$ can only be
            $\beta(\dot{L}'(0))$ and $K_p = |L(0)|$ (*cf.* A.4 and Equation (16)).
        **if** $\iota > \pi = 1$ *or if* $\Delta = 2 \land \dot{L}'(\mathbf{j}^*) \not\equiv \mathbf{y}(K)$ **then**
            $\Delta_p = \Delta$ must hold again and both $K_p$ and $L(\mathbf{j}^*)$ can be retrieved
            non-ambiguously using $\beta$ or $\beta'$ and (17). Then, $\forall j \neq \mathbf{j}^*$,
            $L(j) = \dot{L}'(j)_0 \ldots \dot{L}'(j)_{K_p-1}$ is given by the first $K_p$ letters of $\dot{L}'(j)$.
        **if** $\pi = 1 = \iota$, $\Delta = 2$, *and* $\dot{L}'(\mathbf{j}^*) \equiv \mathbf{y}(K)$ **then**
            $\Delta_p = 1 < \Delta$ and $K_p = K'_{\mathbf{j}^*+1}$ and $L = \bigl(L(0)\bigr)$ where
            $L(0) = \dot{L}'(\mathbf{j}^*+1)_0 \ldots \dot{L}'(\mathbf{j}^*+1)_{K_p}$ is given by the first $K_p$ letters of $\dot{L}'(\mathbf{j}^*+1)$.
    **if** $\pi > 1$ **then**
        $\Delta_p < \Delta$ must hold and there exists a rotation $\rho^q(w)$ of $w$ such that $\dot{L}'^q$ starts
        with a series of identical words. Let $v \equiv w$ be such that the length $D$ of this
        series in $L(v, \Delta) = L''$ is the longest (*cf.* A.5). $D$ satisfies $D \geq \Delta/2$. By definition of
        $\Gamma$ in this case, $\Delta_p = \Delta - D$ must hold, as well as $K_p = \text{lcm}\{K''_j, \ j \geq D\}$ where $K''_j$
        is the primitive period of $L''(j)$. Thus, we let $J \in (\mathbf{W}_1^{K_p})^{\Delta_p}$ be the list s.t.
        $\forall j < \Delta_p$, $J(j) = L''(j+D)_0 \ldots L''(j+D)_{K_p-1}$ is defined by the first $K_p$ letters of
        $L''(j+D)$. Then, $L = J^{-q}$ must hold.

**if** $K = 6$ **then**
    **if** $\Delta > 2$ *and* $\dot{L}'$ *contains one sub-word* $\dot{L}'(\mathbf{j}^*)$ *that has a different period from all the*
    $\Delta - 1$ *other sub-words:* $K_{\mathbf{j}^*} \neq K_j = 6/K_{\mathbf{j}^*}, \forall j \neq \mathbf{j}^*$ **then**
        (*cf.* Item 1 in A.6) $K_p = K_j, \forall j \neq \mathbf{j}^*$, $\Delta_p = \Delta$, and list $L$ can be retrieved from the
        knowledge that in this case, (20) is satisfied by $\dot{L}'$.
    **if** $\dot{L}'$ *contains $D$ identical, consecutive sub-words, where* $\Delta - 1 > D > 2$ **then**
        Let $\mathbf{z}$ be the value of these. Let $K_{\mathbf{z}} \in \{2,3\}$ be its period. Let $K_p = \overline{K}_{\mathbf{z}}$ be the
        common period of the remaining sub-words of $\dot{L}'$, and let $\Delta_p = \Delta - D$. List $L$
        can be retrieved from the knowledge that this case corresponds to Item 2 of
        A.6.
    **else**
        $\dot{L}'$ contains at most two consecutive sub-words of same period and $L$ can be
        retrieved *non-ambiguously* from the knowledge that this case corresponds to
        Items 3 and 4 of A.6. .

**Algorithm 1**: Proof of the injectivity of $\Gamma$.

## B Comparing the behaviours of BACs and BADs: proof of Theorem 2

Lemma 7 below relies on the results given in Table 1 as well as on some properties satisfied by the Lucas [22] and Perrin [1] sequences in relation, respectively, to the two roots of $x^2 - x - 1 = 0$, *i.e.* the **golden ratio** $\mathfrak{g} = \frac{1+\sqrt{5}}{2} \approx 1.61803399$ and $\overline{\mathfrak{g}} = 1 - \mathfrak{g} = \frac{1-\sqrt{5}}{2} \approx -0.61803399$, and to the three roots of $x^3 - x - 1 = 0$ which are the **plastic number** [32] $\pi \approx 1.32471796 \in \mathbb{R}$, $\nu = \frac{1}{2}(-\pi + i \cdot \sqrt{\frac{3}{\pi} - 1})$ and its complex conjugate $\overline{\nu}$.

**Lemma 7.** *For a divisor $p = K_p \Delta_p$ ($\Delta_p = \gcd(\omega, p)$) of the order $\omega$ of a mixed* BAD *and of a negative* BAD*, the numbers of configurations of period $p$ are bounded respectively as follows:*

$$\mathfrak{g}^p \sim \mathrm{X}^{-+} \leq \sqrt{3}^p \quad \text{and} \quad \pi^p \sim \mathrm{X}^{--} \leq \begin{cases} 3^{\frac{p}{3}} & \text{if } K_p = 3 \\ \sqrt{2}^p & \text{if } K_p \neq 3 \end{cases}$$

*where $\Delta$ is the gcd of cycle lengths.*

*Proof.* First, the Lucas sequence satisfies [22]: $\forall n \in \mathbb{N}$, $\mathrm{L}(n) = \mathfrak{g}^n + \overline{\mathfrak{g}}^n = \mathfrak{g}^n + (-\frac{1}{\mathfrak{g}})^n$. Consequences of this and of Table 1 are: $\mathrm{X}^{-+}(p) = \mathrm{L}(K_p)^{\Delta_p} = (\mathfrak{g}^{K_p} + \overline{\mathfrak{g}}^{K_p})^{\Delta_p} \xrightarrow[K_p \to \infty]{} \mathfrak{g}^p$ proving $\mathrm{X}^{-+}(p) \sim \mathfrak{g}^p$. Moreover, using $\mathfrak{g}^2 = 1 + \mathfrak{g}$ and $\overline{\mathfrak{g}} = -\frac{1}{\mathfrak{g}}$ and the binomial formula, we derive:

$$\begin{aligned} \mathrm{X}^{-+}(p) &= \sum_{k \leq \Delta_p} \binom{\Delta_p}{k} (-\mathfrak{g}^2)^{K_p k} \cdot \overline{\mathfrak{g}}^p = \overline{\mathfrak{g}}^p \cdot ((-\mathfrak{g}^2)^{K_p} + 1)^{\Delta_p} \\ &= (-1)^p \cdot |\overline{\mathfrak{g}}|^p \cdot ((-1)^{K_p} \mathfrak{g}^{2K_p} + 1)^{\Delta_p} = \begin{cases} |\overline{\mathfrak{g}}|^p \cdot (\mathfrak{g}^{2K_p} - 1)^{\Delta_p} & \text{if } K_p \text{ is odd} \\ |\overline{\mathfrak{g}}|^p (\mathfrak{g}^{2K_p} + 1)^{\Delta_p} & \text{if } K_p \text{ is even.} \end{cases} \end{aligned} \quad (21)$$

Let us note that if $p$ is odd (and necessarily so are $K_p$ and $\Delta_p$), then $\mathrm{X}^{-+}(p)$ is maximal when $\Delta_p$ is minimal, *i.e.* when $\Delta_p = 1$. And if $p$ is even then, on the contrary, $\mathrm{X}^{-+}(p)$ is maximal when $\Delta_p$ is maximal, *i.e.* when $\Delta_p = p/2$. In both cases:

$$\mathrm{X}^{-+}(p) \leq |\overline{\mathfrak{g}}|^p (\mathfrak{g}^{2K_p} + 1)^{\Delta_p} \leq |\overline{\mathfrak{g}}|^p (\mathfrak{g}^4 + 1)^{\frac{p}{2}} = \frac{(3 + 3\mathfrak{g})^{\frac{p}{2}}}{\mathfrak{g}^p} = 3^{\frac{p}{2}},$$

which proves the first inequality of Lemma 7.

The rest of Lemma 7 derives from Table 1 and from the following relation that is satisfied by the Perrin sequence [1]: $\forall n \geq 2$, $\mathrm{P}(n) = \pi^n + \nu^n + \overline{\nu}^n$. Indeed, this yields $\mathrm{X}^{--}(p) = (\pi^{K_p} + 2\cos(\arg(\nu^{K_p})) \cdot |\nu|^{K_p})^{\Delta_p}$ where $|\nu| = 1/\sqrt{\pi} < 1$, and thus:

$$(\pi^{K_p} - 2|\nu|^{K_p})^{\Delta_p} \leq \mathrm{X}^{--}(p) \leq (\pi^{K_p} + 2|\nu|^{K_p})^{\Delta_p}.$$

Since $(\pi^{K_p} \pm 2|\nu|^{K_p})^{\Delta_p}/\pi^p = \left(1 \pm 2/\pi^{\frac{3}{2}K_p}\right)^{\Delta_p} \xrightarrow[K_p \to \infty]{} 1$, we have: $\left\|\frac{\mathrm{X}^{--}(p)}{\pi^p} - 1\right\| \xrightarrow[p \to \infty]{} 0$.

Now, if $K_p = 3$, then $\mathrm{X}^{--}(p) = \mathrm{P}(3)^{\Delta_p} = 3^{\frac{p}{3}}$. Generally, by the definition of $\pi$, $\forall a \geq$ $\pi$, it holds that $a + 1 \leq a^3$. As a consequence, if, for some $b \in \mathbb{R}$, $P(n) \leq ba^n$, $\forall n \leq m+1$, then: $P(m+3) = P(m+1) + P(m) \leq ba^m(a+1) \leq ba^{m+3}$ and by induction on $m$, $\forall n$, $P(n) \leq ba^n$. Therefore, to prove the last inequality of Lemma 7, it suffices to check that it is satisfied for the base cases of the corresponding induction of this form, where $b = 1$ and $a = \sqrt{2}$.

□

Finally, for any network $\mathcal{N}$ that is either a BAC or a BAD of order $\omega$ (where $\omega = K\Delta$ as before in the case of a BAD), let:

$$\mathtt{a} = \begin{cases} 2 & \text{if } \mathcal{N} \text{ is a positive BAC or BAD} \\ \sqrt{3} & \text{if } \mathcal{N} \text{ is a mixed BAD} \\ \sqrt{2} & \text{if } \mathcal{N} \text{ is a negative BAC or a negative BAD s.t. } K \bmod 3 \neq 0 \\ 3^{1/3} & \text{if } \mathcal{N} \text{ is a negative BAD s.t. } K \bmod 3 = 0. \end{cases} \quad (22)$$

Then, using Table 1, the formulation of $\mathtt{A}(\omega)$ in terms of Dirichlet convolutions (*cf.* Section 6), and Lemma 7 above, the following can be drawn immediately by noting that both $\mathtt{X}$ and the Euler totient $\varphi$ are non-negative:

$$\mathtt{A}(\omega) \leq \frac{(\varphi \star \mathtt{Y})(\omega)}{\omega} \text{ where } \forall p | \omega, \mathtt{Y}(p) = \mathtt{a}^p. \quad (23)$$

This combined with Theorem 1 directly yields Theorem 2. □